\newcommand{\beq}{\begin{quote}}

\newcommand{\enq}{\end{quote}}
\newcommand{\be}{\begin{equation}}
\newcommand{\en}{\end{equation}}
\newcommand{\del}{\delta}

\newcommand{\te}{\theta}
\newcommand{\rh}{\rho}
\newcommand{\ov}{\overline}

\documentclass[prb,preprint]{revtex4-1} 
\usepackage{amsmath}  % needed for \tfrac, \bmatrix, etc.
\usepackage{amsfonts} % needed for bold Greek, Fraktur, and blackboard bold
\usepackage{graphicx} % needed for figures

\begin{document}

\title{Conical tip in frozen water drops}

\author{Michael Nauenberg}
\email{michael@physics.ucsc.edu} % optional
\affiliation{Department of Physics, University of California , Santa Cruz, CA 95064}

\date{\today}

\begin{abstract}
A theory is presented for the formation of a conical tip in water drops
that  are frozen on a flat surface below freezing temperature.   For  the known ice to water density ratio $r=.917$, 
 the angle of aperture of this cone is found to be $\theta=33.55^0$, consistent with observations. 
\end{abstract}

\maketitle 

\begin{figure}[h!]
\centering
\includegraphics[width=12cm]{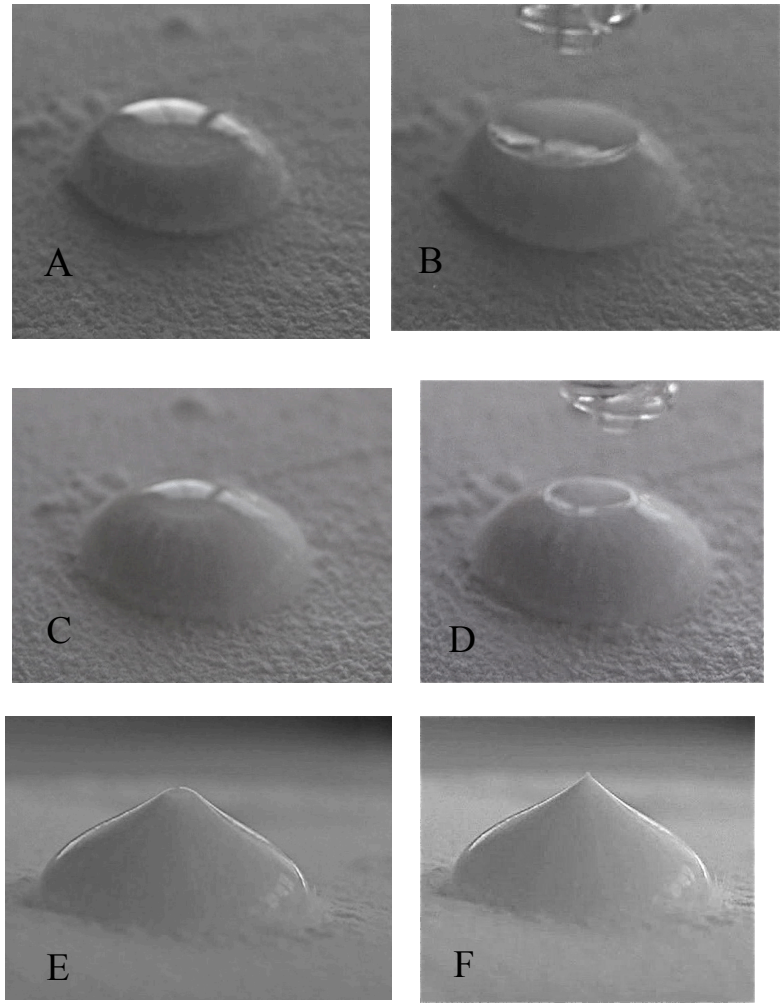}
\caption{Four stages of a water drop freezing on dry ice. A)  Shadow on drop shows the
   boundary of ice-water interface. B) Concave
   shape of  the ice-water  interface, made visible  after the water is removed.  C) and D) Later  stage of the freezing process.  E) Shape of     drop  just before the formation of cusp shown in F.}
\label{}
\end{figure}

When a  small drop of water   is frozen on a flat surface below freezing temperature,  a cusp  appears at the tip of the drop; see Fig1 F.  The origin
of this cusp is attributed to the decrease in density when water freezes,
but a quantitative understanding has not been achieved.  
D.M Anderson et al.  studied 
this solidification process experimentally, and considered an analytic solution  under the assumption that the solidification 
front  is a planar surface \cite{anderson}.  Another  theory for the formation of this tip was developed 
by Snoeijer and Brunet  \cite{snow}, but  they also assumed that this front is flat,
and for the known ice-water density ratio they did not find the  formation of a cusp.   Recently,  however,  I demonstrated  experimentally \cite{michael}that   when this front approaches  the tip of the drop,  its shape becomes  strongly concave; see Figs.1B,D.
This behavior  has  been confirmed independently  by O.R. Enr\'{i}quez et al.  \cite{oscar},  \cite{snow2}. In light of this evidence,  I develop a theory  that leads to  the formation of a conical cusp at the tip of the frozen drop.
  During the freezing process, the surface 
of the front becomes increasingly concave, and the main assumption  in this theory is that near the
end of this process,   the remaining liquid becomes a spherical drop which has  the curvature of this front.   This assumption is justified when the  
liquid phase remaining in the drop becomes 
very small, and then  it might be  expected that surface tension plays a dominant role that would confine it  into  a spherical shape.  

\begin{figure}[h!]
\centering
\includegraphics[width=14cm]{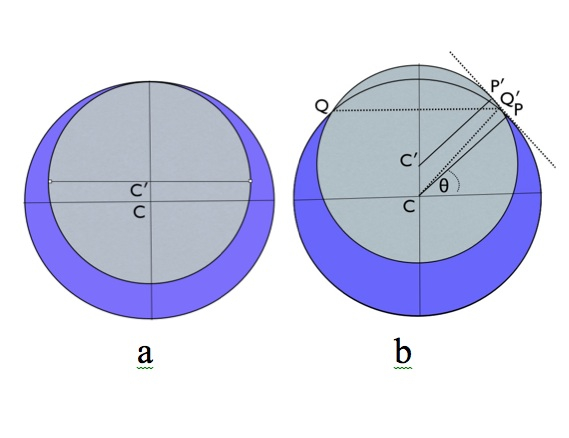}
\caption{Supposing  that  the density of the liquid 
phase (grey), and the ice phase (blue) were equal,  Fig.2 (a)  illustrates the freezing process of a drop during a short time interval.  Fig 2(b) corresponds to the real case that the density of ice is
less than that of water, and the volume of the ice phase increases as shown, pushing the remaining liquid upwards}
\label{}
\end{figure}

The basic idea for this  calculation is illustrated in Fig.2. Supposing that the density of water and ice were equal,  the larger
circle in Fig.2(a)  with center at $C$  and radius $\rho$,  represents a
drop of water, that after a short time interval when a small portion of it  freezes into ice (shown in blue),  
 is represented  by  the smaller circle  with  center at $C'$,  and radius
$\rho-\del \rho$.   Since the freezing process occurs from below, 
the center of the smaller drop is displace vertically by an amount  $\overline{CC'} =\del \rho$.  But actually
the density of ice is less than the density of water,  and therefore the increase in the volume of the frozen liquid implies that the center of
the remaining drop of water is displaced
upwards by an additional amount $\overline{CC'}-\del \rho $ shown in Fig.2(b). The  calculation then proceeds as follows:

Referring to Fig.2(b), let $\overline {PP'}$ be one of the two  tangent lines common to both circles, and $\ov{CP}, \ov{C'P'}$  two  perpendicular lines  to this tangent
from $C$ and $C'$,  at an angle $\te$ from the horizontal.  Then the intersections of the two circles at  the 
points $Q$ and $Q'$ shown in Fig.1(b), lie along the  horizontal
line $\ov{QQ'}$  that intersects the tangent line approximately half way between $P$ and $P'$. Hence, the difference in volume $\del V$ between the segments  of the larger drop  and the smaller one that lies below their  common disk  
with diameter $\ov{QQ'}$,  consist of two parts:  half  the  
volume difference  $2\pi \rho^2 \del \rh$  between the two spheres (to first order in $\del \rh$),
and the corresponding difference  in  the volume of
the segments contained between 
this common disk,   and the disk through the  middle of each sphere. 

Let $x=\rh cos\te$.  The volume of this  segment  for the larger sphere is
\be
\pi \int_0^{\te+\del \te} d\te x^3=\pi\rh^3(sin\te-\frac{1}{3}sin^3\te+\del \te \cos^3\te)
\en
where $\del \te=\del z \cos \te /2 \rh $, and $\del z=\overline{CC'}$ is  related to $\del \rh$ 
by  $\del z =\del \rh /\sin \te$. The volume of the corresponding segment associated
with the smaller  drop is obtained 
by the replacements  $\rh \rightarrow \rh -\del \rh$, and $\del \te \rightarrow -\del \te$. Hence,  the difference of the volumes of these
two segments is 
$ \pi \rh^2 (\del \rh  (3\sin\te-sin^3\te)+2 \rh \del \te cos^3\te)$,
and substituting $2\rh \del \te=\del \rh/\tan\te$, and adding the difference in the volume of half of each of the two spheres,  one obtains
\be
\del V=\pi \rh^2 \del \rh(2 +3\sin\te-sin^3\te+\frac{\cos^4\te}{sin \te}) .
\en
After freezing, the  water initially contained  in  the difference of volume $\del V'=4\pi \rh^2 \del \rho$ between these  two
spherical drops, expands   by a fraction $\xi=1/ r$, where $r$ is the ratio of the density of ice to water.
Hence, setting $\xi  \del V' =\del V$, we obtain a relation for the cone angle $\te$ in terms of $\xi$,
\be
\label{xi1}
\xi=\frac{1}{4}(2 +3\sin\te-sin^3\te+\frac{\cos^4\te}{sin \te})
\en

An alternative method to obtain this relation  is to {\it assume} that the volume of water 
in a spherical drop of radius $\rho$  in contact with the concave ice-water  front eventually
 freezes into a cone  of height $h$  and angle of aperture  $\theta$, where  $h=\rh \cos^2 \te /sin \te$.
The volume of such a cone is 

\be
\label{vc}
V_c=\pi \int_0^{h}dz z^2 \tan^2 \te= \frac{\pi}{3}\rho^3 \frac{\cos^4\te}{ \sin \te}.
\en
The volume $V_s$  of the remaining segment of the drop of radius $\rho$ below this  cone is
\be
\label{vs}
V_s =\frac{2\pi}{3}\rh^3 +\pi \rh^3(\sin \te -\frac{1}{3}sin^3 \te).
\en 
Hence, the water to ice volume expansion  implies that 
\be
\xi V=V_c+V_s,
\en
where $V=(4/3) \pi\rh^3$ is the  initial volume  of the drop, i.e. before freezing.
Then substituting Eqs. \ref{vc} and \ref{vs} for $V_c$ and $V_s$,  one  recovers Eq. \ref{xi1} for $\xi$.
Notice, however, that in  this derivation,  the occurrence of a  tip in the shape of a {\it cone} is an {\it assumption} that was 
previously deduced.

If  the density of water and ice were the same, $\xi=1$,  and Eq. \ref{xi1} yields  $\te=90^0$ indicating, as expected,
that a  tip is not formed.  But for the known  density ratio
 $r=.917$, corresponding to $\xi=1/r=1.0905$,  a conical tip is obtained with aperture $\te=33.55^0$  by
 solving Eq.\ref{xi1}.   This cone aperture is in good agreement  with the frozen drop tip  shown in Fig.1 F.  
 Observations of  such frozen drops   indicate that  the  conical tip is exceedingly   small, implying that  it appears  only at  the very last moment  before the entire drop is frozen.

\end{document}